\documentstyle[12pt]{article}
\begin{document}
\def\thebibliography#1{\section*{REFERENCES\markboth
 {REFERENCES}{REFERENCES}}\list
 {[\arabic{enumi}]}{\settowidth\labelwidth{[#1]}\leftmargin\labelwidth
 \advance\leftmargin\labelsep
 \usecounter{enumi}}
 \def\newblock{\hskip .11em plus .33em minus -.07em}
 \sloppy
 \sfcode`\.=1000\relax}
\let\endthebibliography=\endlist

\hoffset = -1truecm
\voffset = -2truecm


\title{\large\bf
A Microscopic View of Parton $k_T$ Effects In High $p_T$ Processes 
}
\author{
\normalsize\bf
R.K.Shivpuri, B.M.Sodermark and A.N.Mitra* \thanks{e.mail: i) ganmitra@
nde.vsnl.net.in ; ii) anmitra@physics.du.ac.in} \\
\normalsize High Energy Lab, Dept of Physics, Univ of Delhi,
Delhi-110007, India  \\
\normalsize *244 Tagore Park, Delhi-110009, India
}
\date{5 April 2001}
\maketitle

\begin{abstract}
A microscopic mechanism is proposed for understanding the rather large
$k_T$ effects ($<k_T>$ = $1-1.5 GeV/c$, as yet unaccounted for by hard 
QCD), found by the Fermilab E706 Collaboration for 530 and 800 GeV/c 
protons incident on light nuclear targets (mass $A$) like Be. The 
essential idea is that such high incident projectile momenta tend to 
break up the confinement barriers for the quark-partons residing in the 
individual nucleonic constituents of the target nucleus that fall in a
tube-like zone around the projectile's path, so that these particles 
tend to behave as a collection of quark-partons confronting 
the beam. Using simple combinatorial principles, the resultant $<k_T>^2$ 
value works out as $(3A_{eff}-1) \beta^2$, where $A_{eff}$ is the
number of affected nucleons in the tube-like zone, and $\beta$ is a 
scale parameter derived from the basic quark-pair interaction. Using 
the previously found results of a Bethe-Salpeter model (attuned to 
$q{\bar q}$ and $qqq$ spectroscopy), one in which a key ingredient is 
the infrared part of the gluon propagator, giving $\beta^2= 0.068 GeV^2$, 
the desired $<k_T>$ range is reproduced, suggesting the persistence of 
$soft$ QCD effects even at high $p_T$.     

\end{abstract}
                                                    
\newpage


 
\section{ Brief Background}
 
Direct (prompt) photon production at high $p_T$ inclusive processes  
is regarded as a valuable tool for extracting the gluon content 
$G(x)$ [1] of hadrons, due to its sensitivity to the basic
Compton process $qg \rightarrow q\gamma$, involving point-like 
photon-quark coupling. This tool is believed to extend the range of
$x$-values beyond the (limited) $x < 0.25$ [2] found from other data.
The main theoretical ingredients for such analysis are perturbative
QCD for the elementary quark-level process on the one hand, and a
model for the inclusion of the nuclear effects on the other. Either
of these methods has had a long history, dating back to the Seventies. 
In particular, the inclusive cross sections for single high-$p_T$
particles produced in hadron-nucleus scattering have long been known
to show an `anomalous' nuclear dependence growing like $A^\alpha$,
where the value of $\alpha$ is itself a function of $p_T$ [3,4]. This
``Cronin effect'' [3] is believed to be due to multiple (mostly double)
scattering effects, which tend to increase the value of $\alpha$ beyond
unity (linear dependence).      
\par
	The other, more natural ingredient, viz., perturbative QCD, 
has itself been stretched to a considerable extent for an understanding 
of the anomalous nuclear enhancement, by extending the same to include
higher twist effects [5]. More fancy corrections include soft gluon 
resummation [6]; factorization and soft gluon divergences [7]; such
issues and related ones have been discussed in the literature [8]. 
Yet the existence of sharp deviations between the measured inclusive 
direct photon production cross-sections and the predictions of 
perturbative QCD (pQCD), even after the inclusion of next to-leading 
order (NLO) effects, seems to betray some chinks [9] in the 
theoretical armour [10]. To examine this problem in more detail, the 
E706 Group employed their high statistics samples of hard scattering 
data on inclusive $\pi^0$ and direct-photon  production cross sections with 
large $p_T$ values, using beams of 530 and 800 GeV/c protons, as well 
as 515 GeV/c pions, incident on a Be target (A=9) [11], and compared them
with the predictions of NLO pQCD [9-10]. Their results indicate, in company 
with other related investigations [12,13], that the interacting quark-partons 
carry significant initial state transverse momentum ($k_T$), as evidenced
from  the greatly improved fits to the data [11] after a phenomenological 
incorporation of these $k_T$ effects in the NLO calculations.   
\par
	In this brief report, we offer a  microscopic mechanism 
for understanding the rather large $k_T$ effects  as inferred from the 
agreement of the data with the predictions of  NLO-pQCD, after smearing 
the latter with a phenomenological (gaussian) $k_T$ distribution with 
$$<k_T> = 1-1.5 GeV/c$$. 

\section{ A Microscopic View Of $<k_T>$ }
 
The essential idea is the following. At incident projectile momenta 
as high as $\sim 1TeV/c$, the confinement barriers that keep the 
quark-partons residing in the individual nucleonic constituents of the 
target nucleus from mixing with their counterparts from neighbouring 
nucleons, are torn apart in a tube-like zone around the projectile's
path, so that all the nucleons within this zone tend to behave like a 
collection of $3A_{eff}$ quark-partons, where $A_{eff}$ is the number
of nucleons falling with this zone. To estimate $A_{eff}$, note that
if the nucleus is sufficiently small (say $A \leq 10$), almost all 
the constituents will lie within this zone, so that $A_{eff} = A$ in 
this case. On the other hand, if the nucleus is rather big, then many 
nucleons will be outside the path of the projectile, and hence largely 
unaffected by the encounter. One therefore expects a more or less sharp 
cut-off value of $A$ beyond which the size of the nucleus will hardly
matter. While a precise estimate of $A_{eff}$ depends on the radius
of the tube surrounding the projectile's path, and is a difficult
geometrical/dynamical problem, a working value is $A_{eff} \approx 10$
which will be adopted for the calculation to follow. This is perhaps an 
oversimplified description  of the scenario, but hopefully captures the 
essential flavour of the basic mechanism: A huge proliferation in the number 
of scattering centers from the nucleonic to the quark-partonic level 
greatly enhances the scope for the transverse momentum distribution in  
the initial nucleus to influence the inclusive scattering process, 
albeit within the constraints of $A \leq A_{eff}$. This is a  
feature which the NLO pQCD theory [10] seems unable to  capture. Ignoring
double scattering effects [3-4], which are presumably small, the 
mechanism is nevertheless expected to depend on the incident energy per 
nucleon, so that a higher mass number ($A$) would need a proportionately 
larger energy to penetrate the confinement barriers in the constituent 
nucleons, but an incident momentum as high as $~>500 GeV/c$ is probably 
high enough to pulverize the nucleons lying within the projectile's path.   
As will be shown below, with the use of simple combinatorial principles, 
the resultant $<k_T>^2$ value works out as $(3A_{eff}-1)\beta^2$, 
where $\beta$ is a scale parameter incorporating the basic (gluonic)
interaction  of the quark pairs. 

\subsection{ A Non-perturbative Estimate of $\beta^2$ }

Since an estimate of $\beta$ is as yet inaccessible to pQCD, it 
calls for a resort to the non-perturbative QCD regime on which 
unfortunately no consensus exists on the calculational techniques, 
necessitating a certain amount of model building.  To minimise
the uncertainties of the model, its key parameters must be
subjected to several crucial tests bearing on its predictions
as a prior check on its reliability. The model we have chosen to
employ is based on a QCD-motivated Bethe-Salpeter equation [14-17] in 
which the key ingredient is the gluon propagator [14] whose infrared 
part incorporates confinement, and thus represents the soft-QCD regime. 
It is calibrated to both $q{\bar q}$ and $qqq$ spectroscopy [15], as
well as to the hadron form factors [17] within a common dynamical 
framework, and makes a definitive prediction for
$\beta^2$ which controls the quark-parton distribution in a
non-strange baryon with the value $\beta^2 = 0.068 GeV^2$ [15b, 17b]. 
This is the input we propose to take as a means of accounting for the 
high $<k_T>$ value needed for the process under study [11], but its
mode of derivation does not concern us here, and may be found in
the cited references [14-17].

\subsection{ Derivation of $<k_T>^2$ = $(3A_{eff}-1)\beta^2$ } 

To fix our ideas on the essential steps of the calculations, 
we start with the quark-parton distribution in the proton wherein
the two normalized internal Jacobi variables ($\xi, \eta$) are [16]
\begin{equation}\label{1}
  \sqrt{3} \xi = p_1-p_2; \quad 3\eta_ = -2p_3+p_i+p_2  
\end{equation}
which can be further broken up into longitudinal $(\xi,\eta)_\parallel$
and transverse $(\xi,\eta)_\perp$ components. Being interested in  
the transverse part of the distribution here [18], we concentrate only on 
\begin{equation}\label{2}
 \rho_\perp = N_\perp \exp {[-\xi_\perp^2-\eta_\perp^2]/\beta^2} 
\end{equation}
where $N_\perp$ is the normalization attuned to $\int \rho_\perp=1$. 
The mean square transverse momentum for the quark-parton distribution 
may now be defined as
\begin{equation}\label{3}
<\xi_\perp^2 + \eta_\perp^2> = \int d^2\xi_\perp d^2\eta_\perp 
(\xi_\perp^2+\eta_\perp^2) \rho_\perp  
\end{equation}
which works out simply to $2\beta^2$. Its square root which may be 
identified as $<k_T>$ = $ \sqrt{2} \beta$ = $0.37 GeV/c$, is 
unfortunately too small to account for the observed value in Be [5], 
but now we are to take account of the mass number ($A_{eff}$) effect. 
\par
	Now the transverse degrees of freedom in a nucleus
with $3A$ quark-partons are $2 \times (3A_{eff}-1)$, (after subtracting      
unity for the overall c.m. motion). This gives the initial 
transverse quark-parton momentum distribution as
\begin{equation}\label{4}
\rho(\xi_{i\perp}) = N_\perp \exp{[-\sum_{i-1}^{3A-1} \xi_{i\perp}^2]/\beta^2} 
\end{equation}
where the normalization constant is constrained by
\begin{equation}\label{5}
\int \Pi_{i=1}^{3A_{eff}-1} d^2\xi_\perp \rho(\xi_{i\perp}) = 1 
\end{equation}
The mean square transverse momentum now works out as
\begin{eqnarray}\label{6}
<k_T>^2  &=& <\sum_{i=1}^{3A_{eff}-1} \xi_{i\perp}^2> \\  \nonumber 
         &=& \int \Pi_{i=1}^{3A_{eff}-1} d^2 \xi_\perp 
(\sum_{i=1}^{3A_{eff}-1} \xi_{i\perp}^2) \rho(\xi_{i\perp}) \\  \nonumber
         &=& (3A_{eff}-1) \beta^2 
\end{eqnarray}    
 Substitution of $\beta^2 = 0.068 GeV^2 $ [15b,17b] gives for $A_{eff}=9$ [11] 
the value $<k_T>$ = $1.33 GeV/c$, which is well within the range  
 of values considered by the E706 Group [11]. 

\section{Resume And Discussion}

We have tried to offer a microscopic description of $k_T$ 
effects in the rather specific context of direct-photon 
production in proton-Be collisions found by the E706 Group [11].
The assumption underlying the description is that the basic scale 
of the transverse quark-parton distribution is the same as that
obtaining in a $qqq$ hadron (proton). And $this$ in turn is 
determined by the parameters of the $infrared$ part of the 
gluon propagator mediating the quark-pair interaction in the
non-perturbative QCD regime, a logic which points to effects
beyond the NLO pQCD scenario [10]. The other aspect of the 
assumption concerns the $A$-dependence of the effect which
comes about from the fact that at sufficiently high incident 
momenta the projectile is able to ``see'' the quark constituents
that come in the immediate zone surrounding the path of the
projectile through the nucleus, and this greatly enhances the 
source of the $k_T$ distribution to almost three times that of 
the number of ``effective''nucleons.  
\par

\begin{center}
\Large \bf
Table I: Major Expts on Direct Photon Production \\
\large \bf
$k_T$ values for various targets\\
\large
\begin{tabular}{|r|l|c|c|}
\hline
Expt &Beam, Target&$k_T$ expt(GeV/c)&$k_T$ theory (GeV/c) \\
E-706 [11]&p , H&0.7&0.52  \\
E-706 [11]&p , Be&1.2-1.4&1.32 \\
E-629 [22]&p , C&    &1.54  \\
\hline
\end{tabular}
\end{center}

	Table I gives a list of major experiments on various targets,
together with the $k_T$-values which fit the data, including the 
corresponding values according to the proposed mechanism.
For sufficiently small $A$, our formula seems to 
account for the observed value of $<k_T>$ needed to understand the 
$Be$ data [11] for which the full $A$ value is hopefully operative. 
For $pp$ [19] and $p{\bar p}$ [20] reactions, $A_{eff} \approx 2$, 
(taking account of the resultant effect of the target and projectile),
and the best fit value at $k_T \approx 0.7 GeV/c$ for both [11] is not 
much different from the value of $0.52 GeV/c$ predicted by our formula
which indicates a fairly strong dependence on the $A$-value.  
Similar experiments have been carried out for $Cu$, $Be$ and $H$ by the
E-706 group [21]; and for $Be$, $C$ and $Al$ by the E-629 group [22], 
but no clear data on $<k_T>$ are as yet available. Although we  expect 
on the basis of our ansatz that for heavier nuclei like $Cu$ [21] and $Al$ 
[22], the $k_T$ value should $not$ increase proportionately with the 
mass number,  the fuller implications of the ansatz would, in all
probability, need several more targets with considerable 
variations in their $A$-values, as well as a wider range of incident 
energies, before the proposed mechanism can claim a proper test.


\begin{thebibliography}{99}
\bibitem{18}
 In a more recent analysis [17b], it has been found that the
$\beta^2$ value associated with the longitudinal distribution
is somewhat lower than for the transverse distribution, which 
however remains essentially at the value quoted in the text.
\end{thebibliography}
\end{document}